\documentclass[12pt]{article}
\usepackage{epsfig}
\usepackage{amsmath}
\usepackage{hhline}
\usepackage{amssymb}
\usepackage{times}
\usepackage{cite}

\newlength{\dinwidth}
\newlength{\dinmargin}
\begin{document}  
\newcommand{\pom}{{I\!\!P}}
\newcommand{\reg}{{I\!\!R}}
\newcommand{\slowpi}{\pi_{\mathit{slow}}}
\newcommand{\fiidiii}{F_2^{D(3)}}
\newcommand{\fiidiiiarg}{\fiidiii\,(\beta,\,Q^2,\,x)}
\newcommand{\n}{1.19\pm 0.06 (stat.) \pm0.07 (syst.)}
\newcommand{\nz}{1.30\pm 0.08 (stat.)^{+0.08}_{-0.14} (syst.)}
\newcommand{\fiidiiiful}{F_2^{D(4)}\,(\beta,\,Q^2,\,x,\,t)}
\newcommand{\fiipom}{\tilde F_2^D}
\newcommand{\ALPHA}{1.10\pm0.03 (stat.) \pm0.04 (syst.)}
\newcommand{\ALPHAZ}{1.15\pm0.04 (stat.)^{+0.04}_{-0.07} (syst.)}
\newcommand{\fiipomarg}{\fiipom\,(\beta,\,Q^2)}
\newcommand{\pomflux}{f_{\pom / p}}
\newcommand{\nxpom}{1.19\pm 0.06 (stat.) \pm0.07 (syst.)}
\newcommand {\gapprox}
   {\raisebox{-0.7ex}{$\stackrel {\textstyle>}{\sim}$}}
\newcommand {\lapprox}
   {\raisebox{-0.7ex}{$\stackrel {\textstyle<}{\sim}$}}
\def\gsim{\,\lower.25ex\hbox{$\scriptstyle\sim$}\kern-1.30ex%
\raise 0.55ex\hbox{$\scriptstyle >$}\,}
\def\lsim{\,\lower.25ex\hbox{$\scriptstyle\sim$}\kern-1.30ex%
\raise 0.55ex\hbox{$\scriptstyle <$}\,}
\newcommand{\pomfluxarg}{f_{\pom / p}\,(x_\pom)}
\newcommand{\dsf}{\mbox{$F_2^{D(3)}$}}
\newcommand{\dsfva}{\mbox{$F_2^{D(3)}(\beta,Q^2,x_{I\!\!P})$}}
\newcommand{\dsfvb}{\mbox{$F_2^{D(3)}(\beta,Q^2,x)$}}
\newcommand{\dsfpom}{$F_2^{I\!\!P}$}
\newcommand{\gap}{\stackrel{>}{\sim}}
\newcommand{\lap}{\stackrel{<}{\sim}}
\newcommand{\fem}{$F_2^{em}$}
\newcommand{\tsnmp}{$\tilde{\sigma}_{NC}(e^{\mp})$}
\newcommand{\tsnm}{$\tilde{\sigma}_{NC}(e^-)$}
\newcommand{\tsnp}{$\tilde{\sigma}_{NC}(e^+)$}
\newcommand{\st}{$\star$}
\newcommand{\sst}{$\star \star$}
\newcommand{\ssst}{$\star \star \star$}
\newcommand{\sssst}{$\star \star \star \star$}
\newcommand{\tw}{\theta_W}
\newcommand{\sw}{\sin{\theta_W}}
\newcommand{\cw}{\cos{\theta_W}}
\newcommand{\sww}{\sin^2{\theta_W}}
\newcommand{\cww}{\cos^2{\theta_W}}
\newcommand{\trm}{m_{\perp}}
\newcommand{\trp}{p_{\perp}}
\newcommand{\trmm}{m_{\perp}^2}
\newcommand{\trpp}{p_{\perp}^2}
\newcommand{\alp}{\alpha_s}

\newcommand{\alps}{\alpha_s}
\newcommand{\sqrts}{$\sqrt{s}$}
\newcommand{\LO}{$O(\alpha_s^0)$}
\newcommand{\Oa}{$O(\alpha_s)$}
\newcommand{\Oaa}{$O(\alpha_s^2)$}
\newcommand{\PT}{p_{\perp}}
\newcommand{\JPSI}{J/\psi}
\newcommand{\sh}{\hat{s}}
\newcommand{\uh}{\hat{u}}
\newcommand{\MP}{m_{J/\psi}}
\newcommand{\PO}{I\!\!P}
\newcommand{\xbj}{x}
\newcommand{\xpom}{x_{\PO}}
\newcommand{\ttbs}{\char'134}
\newcommand{\xpomlo}{3\times10^{-4}}  
\newcommand{\xpomup}{0.05}  
\newcommand{\dgr}{^\circ}
\newcommand{\pbarnt}{\,\mbox{{\rm pb$^{-1}$}}}
\newcommand{\gev}{\,\mbox{GeV}}
\newcommand{\WBoson}{\mbox{$W$}}
\newcommand{\fbarn}{\,\mbox{{\rm fb}}}
\newcommand{\fbarnt}{\,\mbox{{\rm fb$^{-1}$}}}
%
%
\newcommand{\qsq}{\ensuremath{Q^2}}
\newcommand{\gevsq}{\ensuremath{\mathrm{GeV}^2}}
\newcommand{\et}{\ensuremath{E_t^*} }
\newcommand{\rap}{\ensuremath{\eta^*} }
\newcommand{\gp}{\ensuremath{\gamma^*}p }
\newcommand{\dsiget}{\ensuremath{{\rm d}\sigma_{ep}/{\rm d}E_t^*} }
\newcommand{\dsigrap}{\ensuremath{{\rm d}\sigma_{ep}/{\rm d}\eta^*} }
\def\Journal#1#2#3#4{{#1} {\bf #2} (#3) #4}
\def\NCA{\em Nuovo Cimento}
\def\NIM{\em Nucl. Instrum. Methods}
\def\NIMA{{\em Nucl. Instrum. Methods} {\bf A}}
\def\NPB{{\em Nucl. Phys.}   {\bf B}}
\def\PLB{{\em Phys. Lett.}   {\bf B}}
\def\PRL{\em Phys. Rev. Lett.}
\def\PRD{{\em Phys. Rev.}    {\bf D}}
\def\ZPC{{\em Z. Phys.}      {\bf C}}
\def\EJC{{\em Eur. Phys. J.} {\bf C}}
\def\CPC{\em Comp. Phys. Commun.}
\newcommand{\rfour}{\mbox{$r^{04}_{00}$}}
\newcommand{\rfive}{\mbox{$r^5_{00}$}}
\newcommand{\rfivecomb}{\mbox{$r^5_{00} + 2 r^5_{11}$}}
\newcommand{\ronecomb}{\mbox{$r^1_{00} + 2 r^1_{11}$}}
\newcommand{\tprim}{\mbox{$t^\prime$}}
\newcommand{\rhoprim}{\mbox{$\rho^\prime$}}
\newcommand{\dme}{spin density matrix element}
\newcommand{\dmes}{spin density matrix elements}
\newcommand{\cost}{\mbox{$\cos {\theta^{\ast}}$}}
\newcommand{\rh}{\mbox{$\rho$}}
\newcommand{\mpipi}{\mbox{$m_{\pi^+\pi^-}$}}
\newcommand{\cm}{\mbox{\rm cm}}
\newcommand{\GeV}{\mbox{\rm GeV}}
\newcommand{\GeVx}{\rm GeV}
\newcommand{\gevx}{\rm GeV}
\newcommand{\GeVc}{\rm GeV/c}
\newcommand{\gevc}{\rm GeV/c}
\newcommand{\MeVc}{\rm MeV/c}
\newcommand{\mevc}{\rm MeV/c}
\newcommand{\MeV}{\mbox{\rm MeV}}
\newcommand{\mev}{\mbox{\rm MeV}}
\newcommand{\MeVx}{\mbox{\rm MeV}}
\newcommand{\mevx}{\mbox{\rm MeV}}
\newcommand{\GeVsq}{\mbox{${\rm GeV}^2$}}
\newcommand{\gevsqc}{\mbox{${\rm GeV^2/c^4}$}}
\newcommand{\gevcsq}{\mbox{${\rm GeV/c^2}$}}
\newcommand{\mevcsq}{\mbox{${\rm MeV/c^2}$}}
\newcommand{\GeVsqm}{\mbox{${\rm GeV}^{-2}$}}
\newcommand{\gevsqm}{\mbox{${\rm GeV}^{-2}$}}
\newcommand{\nb}{\mbox{${\rm nb}$}}
\newcommand{\nbinv}{\mbox{${\rm nb^{-1}}$}}
\newcommand{\pbinv}{\mbox{${\rm pb^{-1}}$}}
\newcommand{\mm}{\mbox{$\cdot 10^{-2}$}}
\newcommand{\mmm}{\mbox{$\cdot 10^{-3}$}}
\newcommand{\mmmm}{\mbox{$\cdot 10^{-4}$}}
\newcommand{\degr}{\mbox{$^{\circ}$}}

\begin{titlepage}

\noindent
DESY 02-027  \hfill  ISSN 0418-9833 \\
March 2002


\vspace*{3cm}

\begin{center}
\begin{Large}

{A measurement of the $t$ dependence of the helicity structure \\
  of diffractive \boldmath{$\rho$} meson electroproduction at HERA}
 
\vspace*{1cm}

H1 Collaboration

\end{Large}
\end{center}

\vspace*{3cm}

\begin{abstract}


The helicity structure of the diffractive electroproduction of $\rho$ mesons,
$e + p \rightarrow e + \rho + Y$, is studied 
in a previously unexplored region of large four-momentum transfer
squared at the proton vertex, $t$: 
$0 < t^\prime < 3~{\rm GeV^2}$, where $t^\prime =  |t| - |t|_{min}$.
The data used are
collected with the H1 detector at HERA in the kinematic domain
$2.5 < Q^2 < 60~{\rm GeV^2}$, $40 < W < 120$~GeV.
No $t$ dependence of the $r^{04}_{00}$ spin density matrix element is found.
A significant $t$ dependent helicity non-conservation from the
virtual photon to the \rh\ meson is observed
for the spin density matrix element combinations \rfivecomb\ and 
\ronecomb. 
These $t$ dependences are consistently described by a perturbative
QCD model based on the exchange of two gluons.

\end{abstract}

\vspace{1.5cm}

\begin{center}
\end{center}

\end{titlepage}

%
%

\begin{flushleft}

C.~Adloff$^{33}$,              
V.~Andreev$^{24}$,             
B.~Andrieu$^{27}$,             
T.~Anthonis$^{4}$,             
V.~Arkadov$^{35}$,             
A.~Astvatsatourov$^{35}$,      
A.~Babaev$^{23}$,              
J.~B\"ahr$^{35}$,              
P.~Baranov$^{24}$,             
E.~Barrelet$^{28}$
% ...

\bigskip{\it
% ...
$ ^{4}$ Inter-University Institute for High Energies ULB-VUB, Brussels;
          Universiteit Antwerpen (UIA), Antwerpen; Belgium$^{ c}$ \\
% ...
$ ^{23}$ Institute for Theoretical and Experimental Physics,
          Moscow, Russia$^{ l}$ \\
 $ ^{24}$ Lebedev Physical Institute, Moscow, Russia$^{ e}$ \\
% ...
$ ^{27}$ LPNHE, Ecole Polytechnique, IN2P3-CNRS, Palaiseau, France \\
 $ ^{28}$ LPNHE, Universit\'{e}s Paris VI and VII, IN2P3-CNRS,
          Paris, France \\
% ...
$ ^{33}$ Fachbereich Physik, Bergische Universit\"at Gesamthochschule
          Wuppertal, Wuppertal, Germany \\
% ...
$ ^{35}$ DESY, Zeuthen, Germany \\
% ...
$ ^c$ Supported by FNRS-FWO-Vlaanderen, IISN-IIKW and IWT \\
% ...
$ ^e$ Supported by the Deutsche Forschungsgemeinschaft \\
% ...
$ ^l$ Partially Supported by Russian Foundation
      for Basic Research, grant    no. 00-15-96584 \\
}

\end{flushleft}

\newpage

\section{Introduction}
                               \label{sect:intro}


\noindent
Measurements of exclusive vector meson (VM) production 
in $ep$ scattering at high energy:
\begin{equation}
e + p \rightarrow e + VM + Y \ ,
    \label{eq:VM_prod}
\end{equation}
have led to considerable recent progress towards an understanding of
diffraction in terms of QCD \cite{h1-rho,zeus,h1-phi,jpsi}. The reaction
is induced by a real or virtual photon and 
$Y$ is either a proton (``elastic'' scattering) or a baryonic system 
of mass $M_Y$ which is much lower than the photon--proton centre of mass 
energy $W$ (``proton dissociative'' scattering).
Particularly sensitive tests of QCD models 
are provided by the study of the helicity structure of the interaction 
and its $t$ dependence, 
$t$ being the square of the four-momentum transfer from 
the incident proton to the scattered system $Y$.  
The scope of this paper is to
test diffractive dynamics through the extension of helicity amplitude
extractions for exclusive $\rho^0$ electroproduction to larger values of
$|t|$ than has previously been possible.

Three angles are defined to characterise the electroproduction 
of vector mesons decaying into two charged particles: 
$\Phi$ is the angle between the VM production plane 
(defined as the plane containing the virtual photon and the VM 
directions) and the electron
scattering plane in the ($\gamma^{\star} p$) centre of mass system,
$\theta^\ast$ and $\varphi$ are the polar and
the azimuthal angles, respectively, of the positively charged decay 
particle in the VM rest frame, the quantisation axis being 
taken as the 
direction opposite to that of the outgoing $Y$ system.
In this paper, the distributions of the angles $\Phi$ and 
$\theta^\ast$ are analysed. 

The angular distributions give access to spin density matrix elements, 
which are bilinear combinations of the helicity amplitudes
$T_{\lambda_{V \! M}\lambda_{\gamma}}$, where $\lambda_{V \! M}$
($\lambda_{\gamma}$) is the vector meson (virtual photon)
helicity~\cite{sch-w}.
In the case of vector meson electroproduction by unpolarised beams 
and their subsequent decay into two pseudoscalar particles 
($\rho \rightarrow \pi^+ 
\pi^-, \phi \rightarrow K^+ K^-$), the $\Phi$ and $\theta^\ast$ 
distributions, integrated over the other two angles, are related to 5 
of the 15 spin density matrix elements $r^k_{ij}$ and 
$r^{kl}_{ij}$ through the relations~\cite{sch-w}
\begin{equation}
  \frac {\rm{d} \sigma} {\rm{d} \Phi} \propto 
   1 +
   \sqrt {2 \epsilon (1+\epsilon)} \ \cos {\Phi} \ (\rfivecomb)
   - \epsilon \ \cos {2 \Phi} \ (\ronecomb)
                                \label{eq:Phi}
\end{equation}
\begin{equation}
  \frac {\rm{d} \sigma} {\rm{d} \cost}  \propto 
   1 - \rfour\ + (3 \ \rfour - 1) \cos^2{\theta^{\ast}} \ ,
                                \label{eq:cost}
\end{equation}
where $\epsilon$ is the polarisation parameter, i.e. the ratio of the 
longitudinal to transverse virtual photon fluxes. For this analysis
$\epsilon \simeq 0.99$~\cite{h1-rho}.

Assuming natural parity exchange to hold 
(
$T_{-\lambda_{V \! M}-\lambda_{\gamma}} = 
 (-1)^{\lambda_{V \! M} - \lambda_{\gamma}} \
        T_{\lambda_{V \! M}\lambda_{\gamma}}$),
these five \dmes\ are related to the five independent complex helicity 
amplitudes by the following relations:
\begin{eqnarray}
 & r^{04}_{00} & \propto \frac {1} {N} \ (|T_{00}|^2 + |T_{01}|^2)            
\nonumber \\
  & r^5_{00} & \propto \frac {1} {N} \ {\rm Re} \ (T_{00} \ 
T_{01}^\dagger) \nonumber \\
  & r^5_{11} & \propto \frac {1} {N} \ ({\rm Re} \ (T_{10} 
T_{11}^\dagger)   
                  - {\rm Re} \ (T_{10} T_{1-1}^\dagger))                
\nonumber \\
  & r^1_{00} & \propto  \frac {-1} {N} \  |T_{01}|^2                       
\nonumber \\
  & r^1_{11} & \propto  \frac {1} {N} \ (T_{1-1} T_{11}^\dagger + 
T_{11} T_{1-1}^\dagger) \nonumber \\
  & {\rm with} & \ \  N =  |T_{00}|^2 \ + \ |T_{11}|^2 \ +  \ 
|T_{01}|^2 \ +
\ 2 \ |T_{10}|^2 \ + \ |T_{1-1}|^2 \ .
%
                                \label{eq:ampli}
\end{eqnarray}
In the case of {\it s}-channel helicity conservation (SCHC),
$\lambda_{VM}=\lambda_{\gamma}$, only the $T_{00}$ and $T_{11}$
helicity ``non-flip''
amplitudes are non-zero,  
$r^5_{00}=r^5_{11}=r^1_{00}=r^1_{11}=0$
and the ratio $R$ of the cross sections for longitudinal to transverse photons
is given by 
$R = 1 / \epsilon \ \rfour / (1 - \rfour ) $.

In recent years, the spin density matrix elements 
describing process~(\ref{eq:VM_prod}) have been measured 
for the elastic electroproduction of $\rho$ and $\phi$ mesons 
in the kinematic range 
$Q^2 > 2.5$~\gevsq\ and 
$|t| < 0.5$~\gevsq~\cite{h1-rho,zeus,h1-phi}, 
\qsq\ being the negative square of the virtual photon four-momentum.
Three main features have emerged from these measurements:
\begin{itemize}
\item the dominance of the longitudinal $T_{00}$ over the transverse
$T_{11}$ helicity non-flip amplitudes;
\item the presence of a small but significant violation of SCHC, 
observed through the non-zero value of the \rfive\ matrix element, 
in which the dominant helicity single flip amplitude
describes the transition from a transverse 
photon to a longitudinal vector meson ($T_{01}$);
\item values compatible with zero for the other amplitudes describing
single ($T_{10}$) or double helicity flip ($T_{1-1}$). 
\end{itemize}

These features are in agreement with calculations based on 
perturbative QCD (pQCD)~\cite{royen,niko,ivanov}.
In these approaches, vector meson electroproduction is described
in the proton rest frame
as the convolution of a virtual photon fluctuation into a 
$q \bar{q}$ pair at a long distance from the target,
a hard interaction mediated by the exchange of two gluons
(each of them must carry sufficiently large transverse momentum 
to resolve the $q \bar{q}$ pair and the proton structure),
and the subsequent recombination of the quark pair into 
a vector meson. 
For massless quarks, the helicity of the $q \bar{q}$ pair is zero, 
such that the helicity of the virtual photon is transferred into 
the projection of the orbital angular momentum of the $q \bar{q}$ 
pair onto the $\gamma^{\star}$ direction.
During the interaction, the helicity and the impact parameter of the 
quark pair are unchanged, but the orbital angular momentum  
can be modified through the transfer of the transverse momentum 
carried by the gluons. 
The helicity of the outgoing vector meson can thus 
be different from that of the incoming photon.
Calculations show that such a helicity flip between the photon and 
the vector meson requires an asymmetric sharing 
of the photon longitudinal momentum by the quark and the 
antiquark~\cite{royen,niko,ivanov}.

For \qsq\ above a few~\gevsq\ and $|t| \lsim\ \qsq $,
the following features are expected 
for the (predominantly imaginary) amplitudes:
\begin{itemize}
\item a ratio constant with $t$ for the helicity conserving amplitudes 
$|T_{11}| \ / \ |T_{00}|$;
\item a $\sqrt {|t|}$ dependence for the ratio of the single helicity
flip to the non-flip amplitudes
$|T_{01}| \ / \ |T_{00}|$ and $|T_{10}| \ / \ |T_{00}|$;
\item a dependence linear with $t$ for the ratio of the double flip to
the non-flip amplitudes
$|T_{1-1}| \ / \ |T_{00}|$;
\item the hierarchy: 
\begin{equation}
|T_{00}| > |T_{11}| > |T_{01}| > |T_{10}| > |T_{1-1}| \ . 
                                \label{eq:hierarchy}
\end{equation}
\end{itemize}
%
These features are expected to hold for proton dissociative 
as well as for elastic scattering.

Compared to previous results~\cite{h1-rho,zeus,h1-phi}, 
the present paper extends 
considerably the $t$ range of the measurement of \dmes\ 
for $\rho$ meson diffractive electroproduction
\begin{equation}
e + p \rightarrow e + \rho + Y \ ; \ \ \ \  
\rho \rightarrow \pi^+ \pi^- \ ,
                                \label{eq:rho_prod}
\end{equation}
where the $\rho$ mass range is defined by restricting the invariant mass 
$M_{\pi \pi}$ of the decay pions to the interval
\begin{equation}
0.6 < M_{\pi \pi} < 1.1 \ {\rm GeV} \ .
   \label{eq:rho_mass}
\end{equation}
%

Elastic and proton dissociative data are combined, and the kinematic 
domain of the measurement is:
\begin{eqnarray}
2.5  < &Q^2& < 60\ {\rm GeV^2}       \nonumber \\
 40 < &W& < 120\ {\rm GeV}           \nonumber \\
 0 < &t'& < 3\ {\rm GeV^{2}} \ .          
    \label{eq:kin_range}
\end{eqnarray}
The variable $t^\prime = |t| - |t|_{min}$ is used for the analysis, 
where $|t|_{min}$ is the minimal value of $|t|$ kinematically required 
for the vector meson and the system $Y$ to acquire their effective 
mass through longitudinal momentum transfer.
The \tprim\ variable, which is very well approximated as the square 
of the transverse momentum of the scattered system $Y$, describes 
the transverse momentum transfer to the target and is thus the relevant 
dynamical variable.
In the  elastic case and for moderate \qsq , $|t|_{min}$ is negligible
and $t^\prime \simeq |t| $.

The large \tprim\ domain covered by the present data 
allows for the first time a detailed study 
of the \tprim\ dependence of the helicity structure of diffractive vector 
meson electroproduction.


\section{Experimental procedure}
                                  \label{sect:expproc}

\subsection{Event selection, kinematic variables 
              and Monte Carlo simulations}
                               \label{sect:selection}

The data used for the present analysis were taken with the H1 detector
in 1997. 
The energies of the HERA proton and positron beams\footnote{
In the following, the word electron will be used for both 
electrons and positrons.} 
were 820 GeV and 27.5 GeV, respectively. 
The integrated luminosity used for
the analysis amounts to 6.0 ${\rm pb^{-1}}$.
The relevant parts of the detector, for which more details 
can be found in~\cite{h1-rho,nim}, are
the central tracking detector, the liquid argon (LAr) and the
backward electromagnetic (SPACAL) calorimeters and the 
forward detectors, which are sensitive to energy flow close to the 
outgoing proton direction,\footnote{
In the H1 convention, the $z$ axis is defined by the colliding
beams, the forward direction being that of the outgoing
proton beam ($z > 0$) and the backward direction that of the electron
beam ($z < 0$).}
i.e. the proton remnant tagger (PRT) and the forward muon detector 
(FMD).

Events corresponding to reaction~(\ref{eq:rho_prod}), in the 
kinematic range defined by relations~(\ref{eq:kin_range}),
are selected by requesting the reconstruction of a cluster 
in the SPACAL calorimeter 
with energy larger than 17 GeV (the scattered electron candidate)
and the reconstruction
in the central tracking detector of the 
trajectories of exactly two charged particles (pion candidates)
with opposite charges, transverse momenta larger than 0.1 GeV and 
polar angles 
confined within the interval $20^{\rm o} < \theta < 160^{\rm o}$. 
To reduce the background due to diffractive production of $\phi$ mesons, 
events 
with $M_{KK}~<~1.04~{\rm GeV}$ are discarded, where $M_{KK}$ is
the invariant mass of the two hadron candidates when 
considered as kaons (no direct hadron identification is performed for 
this analysis). 
In order to reduce both QED radiative corrections and background
contributions in which there are unreconstructed particles, a cut 
$ E-p_z > 52\ {\rm GeV}$ is applied. $E-p_z$ is the difference 
of the energies and the longitudinal momenta of the scattered electron 
(measured in the SPACAL) and the pion candidates (measured in the 
central tracking detector); it is expected to be close to twice 
the incident electron beam energy, i.e. 55 GeV, if no other particles 
have been produced except for the forward going system $Y$.
To avoid backgrounds due to the diffractive production
of systems decaying into two charged and additional neutral particles, 
all events are rejected in which a cluster, which is not associated
with the electron or the two charged pion candidates, is reconstructed 
with polar angle larger than $20^{\rm o}$ and energy larger than 
400~MeV (300~MeV) in the LAr (SPACAL) calorimeter.

The \rh\ three-momentum is computed as the sum of the two charged 
pion candidate momenta.
The variable \qsq\ is reconstructed using the double angle 
method~\cite{da}:
\begin{equation}
Q^2 = \frac {4 E_0^2 \ \sin {\theta_{\rho}} \ (1+\cos{\theta_e})}
{\sin{\theta_e} + \sin{\theta_{\rho}} - 
\sin{(\theta_e+\theta_{\rho})}} \ ,
                                \label{eq:qsq}
\end{equation}
where ${E_0}$ is the energy of the incoming electron and 
$\theta_e$ and $\theta_{\rho}$ are the scattered electron and \rh\ 
meson polar angles, respectively.
The variable $W$ is calculated using the Jacquet-Blondel 
method~\cite{jb}:
\begin{equation}
W^2 = y \cdot s - \frac {{p_{t, \rho}}^2} {1-y} \ \ , \ \ {\rm with} \ \
  y= \frac{E_{\rho} - p_{z, \rho}} {2 E_0} \ ,
                                \label{eq:w}
\end{equation}
$s$ being the square of the $ep$ centre of mass energy and $E_{\rho}$,
$p_{z, \rho}$ and $p_{t, \rho}$ being the energy, the longitudinal and
the transverse momentum of the \rh\ meson, respectively. 
The electron transverse momentum is computed as
\begin{equation}
 p_{t, e} = \frac {2  E_0 - E_{\rho} + p_{z, \rho}}
               {\tan (\theta_e / 2)} \ .
                                \label{eq:pte}
\end{equation}
The variable \tprim\ is then determined 
from the scattered electron and 
\rh\ momentum components transverse to the beam direction as
\begin{equation}
t^\prime \simeq (\vec{p}_{t, miss})^2 = 
  (\vec{p}_{t, e} + \vec{p}_{t, \rho})^2 \ . 
                                \label{eq:tprim}
\end{equation}
%


The selected events are classified in two categories, corresponding 
to the absence or presence of activity in the forward part 
of the H1 detector.
An event is classified in the ``notag'' sample when no 
signal above noise is detected in the PRT and the 
FMD, and no track and no LAr cluster with energy larger than
400 MeV is
reconstructed with polar angle $\theta < 20^{\rm o}$. 
Conversely, events are classified in the ``tag'' sample if a 
signal is observed in either the PRT or the FMD, or if a track or 
a LAr cluster with energy larger than 400 MeV is reconstructed 
in the forward part of the H1 detector 
($\theta < 20^{\rm o}$).\footnote{
In the case of proton dissociative 
scattering, this corresponds to an excitation mass of the target $M_Y 
\lsim\ 25$\gev .}
For the tagged events, a pseudorapidity interval of at least 2.2 
units\footnote{
The pseudorapidity $\eta$ of an object with polar angle 
$\theta$ is defined as
$\eta = - \log \ \tan (\theta / 2)$.}
is required between the most backward track or LAr cluster with 
$\theta < 20^{\rm o}$ (or the forward edge of the LAr calorimeter in 
the absence of track or cluster with $\theta < 20^{\rm o}$), and 
the most forward charged pion candidate.
To first approximation, the notag and tag samples could be attributed
to the elastic and the proton dissociative processes, respectively. 
However, elastic events fall in the tag sample when $|t|$ is 
large enough for the scattered proton to hit the beam pipe walls 
or adjacent material, leading to secondary particles which give 
a signal in the forward detectors. 
This effect becomes significant for $\tprim \ \gapprox \ 0.75$~\gevsq .
Conversely, proton dissociative events are classified in the notag 
sample for small masses, $M_Y \ \lapprox \ 1.6$~GeV,
or in the case of inefficiencies of the forward detectors.

\begin{figure}[htbp]
\vspace{-0.cm}
\begin{center}
\epsfig{file=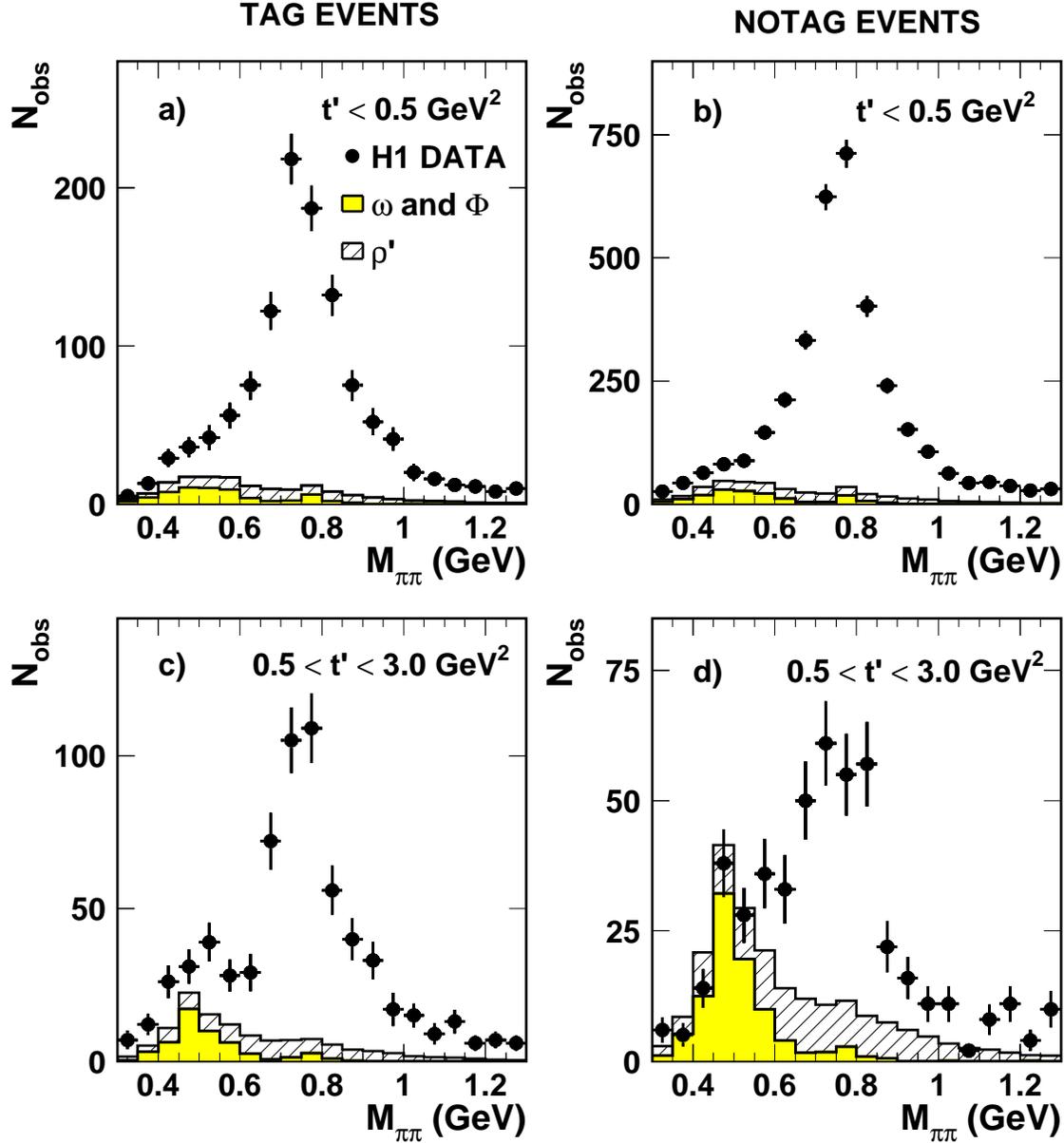,width=16.cm}
\end{center}
 \caption{Uncorrected $\pi^+\pi^-$ mass distributions
 for the selected events with  $0.3 < M_{\pi \pi} <~1.3$~GeV.
 The shaded areas describe the $\omega$ and $\phi$ backgrounds and
 the hatched areas the \protect\rhoprim\ background.
 a) and c) show tag events with \tprim $<$ 0.5 \gevsq\ 
 and 0.5 $<$ \tprim  $< 3.0 \ {\rm GeV^{2}}$, respectively;
 b) and d) show notag events.}
 \label{fig:mass}
 \end{figure}

The uncorrected $\pi^+\pi^-$ mass distributions are shown over the 
extended mass region $0.3 < M_{\pi \pi} < 1.3$~GeV
in Fig.~\ref{fig:mass}, separately for the tag and the notag samples, 
for $ \tprim  < \ 0.5 \ \rm{GeV}^2$ and for 
$0.5 < \tprim < 3 \ \rm{GeV}^2$. 
Clear \rh\ meson signals are visible in all distributions.


Monte Carlo simulations based on the DIFFVM program~\cite{diffvm} 
including QED radiation~\cite{heracles} are used to describe the  
production and decay of $\rho$ vector mesons in elastic and proton
dissociative scattering, and to correct the data for acceptance, 
smearing and radiative effects.
The simulations include the angular distributions 
corresponding to the measurements of the present analysis for
the \rfour\ matrix element (\cost\ distribution) and the
\rfivecomb\ and \ronecomb\ combinations ($\Phi$ distribution).
Other angular distributions and correlations are taken in the 
SCHC approximation, and the 
$\cos \delta$ parameter, which describes the interference
between the longitudinal and transverse amplitudes, is taken
from the elastic 
scattering measurement~\cite{h1-rho} in the relevant \qsq\ range.
The exponential slope of the $t$ distribution is  
$b_{el} = 7$~\gevsqm\ for elastic scattering~\cite{h1-rho} and 
$b_{pd} = 1.7$~\gevsqm\ for proton dissociative scattering, values 
which describe well the \tprim\ distribution of the present data.
The \tprim-integrated cross section ratio for proton dissociative
to elastic scattering is taken as 0.75 in the present $Q^2$ 
range~\cite{h1-pd}.
For proton dissociative scattering, the $M_Y$ spectrum is parameterised
as ${\rm d} \sigma / {\rm d} M_Y^2 \propto 1/M_Y^{2.15}$ 
(see~\cite{goulianos}) and corrections are applied for the loss 
of events with large $M_Y$ values when particles of the dissociation 
system are reconstructed in the detector with polar angles 
$\theta > 20^{\rm o}$ .
All these parameters have been varied in the simulation as a part of the
systematic error analysis.

DIFFVM simulations have also been used for $\omega$, $\phi$ and
\rhoprim\ background studies (see next section).
In all cases, the $t$ slopes are chosen to be  $b_{el} = 6$~\gevsqm\
for elastic scattering and $b_{pd} = 2.5$~\gevsqm\ for proton
dissociative scattering. The ratios of the proton dissociative to
elastic channels, integrated over \tprim, are 0.75.
In the absence of measurements in electroproduction, the angular 
distributions for $\omega$, $\phi$ 
(except for $\phi \rightarrow K^+ K^-$~\cite{h1-phi}) 
and \rhoprim\ are treated as isotropic.


\subsection{\boldmath{$\omega$}, \boldmath{$\phi$} and 
     \boldmath{\rhoprim } backgrounds}
                               \label{sect:bg}


Diffractive electroproduction of $\omega$ and $\phi$ mesons can 
fake \rh\ production through the decay channels
\begin{eqnarray}
&& \omega \rightarrow \pi^+ \pi^- \pi^0 \ , \nonumber \\    
&& \phi \rightarrow \pi^+ \pi^- \pi^0 \ , \ \ \ \ \ \phi \rightarrow 
         K^0_S K^0_L \ ,     
                                \label{eq:omegaphi}
\end{eqnarray}
if the decay photons of the $\pi^0$ or the $K^0_L$ meson are not 
detected. 
This happens if the deposited energy is associated with the charged 
pion tracks or does not pass the detection threshold in the detector.
The $p_t$ imbalance of the event due to the loss of 
particles can then be interpreted as \rh\ production at large \tprim , 
following eq.~(\ref{eq:tprim}).
These background contributions, which are concentrated below 
the selected \rh\ mass range~(\ref{eq:rho_mass}), are estimated using 
the Monte-Carlo simulations.
The ratios of the production cross sections $\omega$~/~$\rho$ 
and $\phi$~/~$\rho$ are, for the present \qsq\ range, taken as
0.09~\cite{zeus-omega} and 0.20~\cite{h1-phi}, respectively.

Another background reaction, particularly important for large \tprim\
in the selected mass range~(\ref{eq:rho_mass})  
is the electroproduction of \rhoprim\ mesons\footnote
{The detailed structure~\cite{pdg} of the states described in the past 
as the \rhoprim(1600) meson is not relevant for the present study.
The name \rhoprim\ is used to imply all vector
meson states with mass in the range 1300-1700 MeV.
In the simulations, the \rhoprim\ mass and width are taken as 1450 MeV 
and 300 MeV, respectively.}
decaying into two charged pions and two $\pi^0$:
%
\begin{equation}
\rho^\prime \rightarrow \rho^+ \pi^- \pi^0 \ , \ \ \ \ \     
  \rho^+ \rightarrow \pi^+ \pi^0       \ \ \ \ \       (+ \ c. c.) \ .
                                \label{eq:rhoprim}
\end{equation}
Again, the non-detection of the two $\pi^0$ mesons induces $p_t$ 
imbalance which fakes \rh\ production at large \tprim .
No measurements exist in the relevant $Q^2$ range of 
the $\rho^\prime \rightarrow \pi^+ \pi^- \pi^0 \pi^0$ to \rh\
cross section ratio. 
The \rhoprim\ contribution is thus determined in 
section~\ref{sect:rhoprim_bg} from the data themselves, 
using the events with $0.5 < \tprim < 3 \ {\rm GeV^2}$.

It is important to recognise that the presence of backgrounds 
at large \tprim\ values affects differently the tag and the notag 
samples defined in section~\ref{sect:selection}.
As mentioned there, genuine production of \rh\ mesons at large \tprim , 
either due to proton dissociative or elastic scattering, usually gives 
a signal in the forward detectors and contributes mainly to the tag 
sample.
In contrast, $\omega$, $\phi$ and \rhoprim\ background events,
produced mainly at low \tprim\ but faking high \tprim\ \rh\ production, 
contribute to either the tag or the notag sample, depending on whether 
the proton dissociates and on the detector response. 
The ratio of the \rh\ signal to background at high \tprim\ is thus 
significantly higher in the tag sample than in the notag sample.


\subsection{Determination of the \boldmath{\rhoprim } 
            background}
                               \label{sect:rhoprim_bg}

\begin{figure}[htbp]
\vspace{-0.cm}
\begin{center}
\epsfig{file=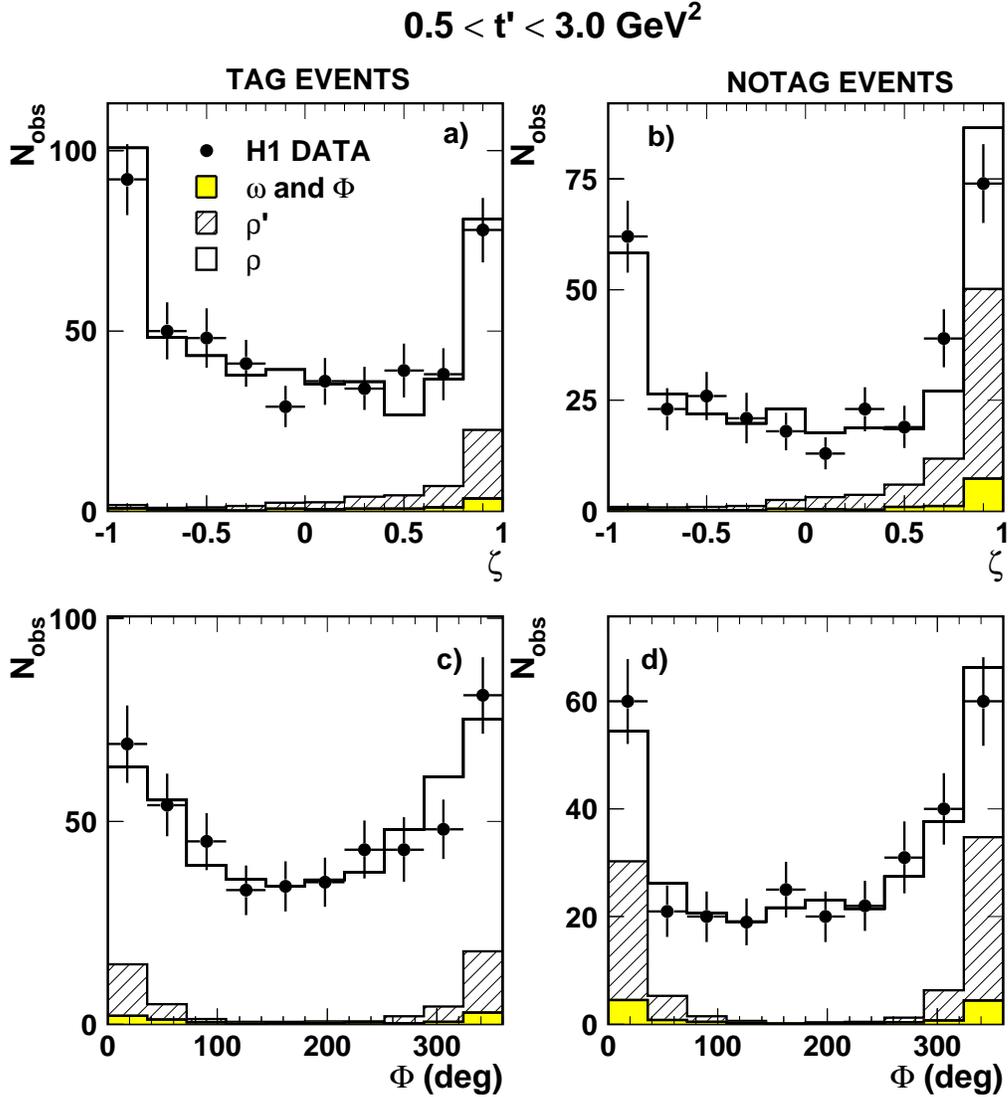,width=14.cm}
\end{center}
\vspace{-1.cm}
\caption{Distributions of a-b) the $\zeta$ and c-d) the $\Phi$
 variables of the selected events in the mass 
 range~(\protect\ref{eq:rho_mass}) with 
 $0.5 < \protect\tprim < 3 \ {\rm GeV^2}$, in the tag (a and c) and 
 in the notag (b and d) sample.
 The shaded areas describe the $\omega$ and $\phi$ background
 as obtained from the simulations. The hatched areas 
 correspond to the \protect\rhoprim\ background and the open areas 
 to the \protect\rh\ contribution, as determined using the iterative 
 fitting procedure described in section~\protect\ref{sect:rhoprim_bg}.}
\label{fig:zeta}
\end{figure}

In order to determine the \rhoprim\ background, a new variable, 
$\zeta$, is introduced:
\begin{equation}
\zeta = \frac {\vec{p}_{t, miss} \cdot \vec{p}_{t, \rho}}
              {|\vec{p}_{t, miss}| |\vec{p}_{t, \rho}|}  \ .
                                \label{eq:y}
\end{equation}
For \rhoprim\ events produced at low \tprim\ and faking high \tprim\ 
\rh\ production, 
$\vec{p}_{t, miss}$ is due to the two missing $\pi^0$ mesons 
and, in the present $Q^2$ range, is generally aligned along the 
($\pi^+, \pi^-$) direction.
This gives for the \rhoprim\ background a $\zeta$ distribution 
peaking around $+1$, as shown in Figs.~\ref{fig:zeta}a-b;
the same effect is found for $\omega$ and $\phi$ production with the 
decay channels~(\ref{eq:omegaphi}).
In contrast, for genuine high \tprim\ $\rho$ production, 
$\vec{p}_{t, miss}$ is the transverse momentum of the scattered 
proton or baryonic system, leading to a flatter $\zeta$ distribution, 
with maxima at $-1$ and $+1$.
The $\zeta$ and the $\Phi$ distributions are strongly
correlated: positive $\zeta$ values correspond to $\Phi$ 
angles close to $0^{\rm o}$ and $360^{\rm o}$, whereas for negative 
$\zeta$, central $\Phi$ values are selected.
This is visible in Figs.~\ref{fig:zeta}c-d, which compare 
background events (for which $\zeta$ is mostly positive) to \rh\ 
events (for which negative $\zeta$ values dominate).


An iterative fitting procedure is performed to estimate the
\rhoprim\ background, whilst dealing with this $\zeta - \Phi$
correlation.
After subtraction of the $\omega$ and $\phi$ backgrounds using
the Monte Carlo simulations, the selected events in the mass 
range~(\ref{eq:rho_mass}) are divided into four subsamples: 
tag events and notag events, separately 
with $\zeta < 0$ and with $\zeta > 0$.
Each of these four samples contains two contributions, due to genuine 
\rh\ production and to \rhoprim\ background.
These eight contributions (eight unknowns) are determined through
an overconstrained fit, performed using the MINUIT 
package~\cite{minuit}, to the numbers of events in the four samples 
(four measurements), under the following six constraints, obtained 
using the Monte Carlo simulations:
%
\begin{enumerate}
\item[i.]
two constraints describe the asymmetry of the $\zeta$
distribution of \rh\ events (i.e. the ratio of the numbers of events 
with $\zeta < 0$ and with $\zeta > 0$), separately for the tag and 
for the notag sample (see the open areas in 
Figs.~\ref{fig:zeta}a and b, respectively); 
\item[ii.]
similarly, two constraints describe the asymmetry of the 
$\zeta$ distribution for \rhoprim\ events, separately for the tag and 
notag samples (hatched areas); 
\item[iii.] 
the last two constraints, defined separately for $\zeta < 0$ and 
for $\zeta > 0$, are the probabilities for any \rhoprim\ event 
(elastic or proton dissociative) to be tagged; 
the ratio of the proton dissociative to elastic \rhoprim\ production 
cross sections is taken as $0.75$ in the simulation.
\end{enumerate}

An estimate of the \rhoprim\ background is thus obtained, and its 
$\Phi$ distribution is computed using the simulation.
Relation~(\ref{eq:Phi}) is then fitted to the $\Phi$ distribution 
in the tag sample, fully corrected for
background, acceptance, smearing and radiative effects, 
to extract values of the \dme\ combinations 
\rfivecomb\ and \ronecomb .
These values are fed back into the \rh\ simulation, leading to a 
modification of the simulated $\Phi$ and hence $\zeta$ distributions,
which provides new values for the constraints describing 
the asymmetry of the latter (see item~i. above).
The fitting procedure is repeated, and the iterative process converges 
after a few steps to stable background estimates, independent of 
the starting values of the spin density matrix elements in the \rh\ 
Monte Carlo simulation.

Fig.~\ref{fig:zeta} presents the $\zeta$ and $\Phi$ distributions
of the selected events with $0.5 < \protect\tprim < 3 \ {\rm GeV^2}$, 
separately for the tag and the notag samples.
They are well described by the superposition of the 
$\omega$ and $\phi$ background, 
the \rhoprim\ background 
and the \rh\ contribution, as determined from 
the iterative fitting procedure. 
The dominant background is found to be from \rhoprim\ production and,
as expected, the backgrounds are larger in the notag sample and 
affect mainly the $\zeta \ > \ 0$ region.

This procedure thus provides an estimate of the $\rho^\prime/\rho$
cross section ratio for $0.5 < \tprim <
3 \ {\rm GeV^{2}} $.
This estimate is extended to $\tprim < 0.5$~\gevsq ,  under the
assumptions quoted in section~\ref{sect:selection} for the $t$ slopes
and for the proton dissociative to elastic cross section ratio.

The background contributions are shown in Figs.~\ref{fig:mass}a-d.
After background subtraction, relativistic Breit-Wigner functions, 
with the Ross-Stodolsky skewing parameter~\cite{rs} left free, 
are fitted to the fully corrected data,
yielding \rh\ mass and width values in 
excellent agreement with expectations~\cite{pdg} and good 
$\chi^2$ values.
The data are thus very well described 
by diffractive \rh\ production with contributions of additional 
$\omega$, $\phi$ and \rhoprim\ backgrounds.
The elastic cross section at low $t$ agrees with previous 
measurements~\cite{h1-rho}.


\subsection{Systematic errors}
                               \label{sect:syst}

In addition to the effect of varying the number of bins, the 
systematic uncertainties affecting the measurements described in
section~\ref{sect:tdepend} are grouped into three classes:

\begin{itemize}
\item {\bf Uncertainties in the amount and shape of the backgrounds} \\
%
The amount of $\omega$ and $\phi$ backgrounds is
varied by $\pm$~50~\%.
The uncertainty in the amount of \rhoprim\ background for 
$ \tprim  > \ 0.5 \ \rm{GeV}^2$
is estimated by varying, for the fit procedure, the $\zeta$
separation between the samples (at $-0.4$, $-0.2$, $0.2$ and $0.4$ 
instead of 0).
The fractions of background events with $\zeta < 0$ (see
Figs.~\ref{fig:zeta}a-b) are also multiplied by 2 and 0.5.
For $ \tprim  < \ 0.5 \ \rm{GeV}^2$, the
$\rho^\prime/\rho$ cross section ratio is changed by $\pm$ 50 \%.
The shape of the background in the $\Phi$ distribution 
(section ~\ref{sect:combin}) is modified by keeping the total amount
fixed, but changing the fraction assigned to the two extreme $\Phi$
bins by $\pm$ 50~\% (see Figs.~\ref{fig:zeta}c-d).
For the \cost\ distribution (section ~\ref{sect:rfour}), the shape of the
background is varied from flat to the same
distribution as that of the \rh\ signal.
The following model uncertainties in the background simulations are also 
included:
the $t$ slopes are varied ($b_{el} = 6 \pm 1 \ \rm{GeV}^{-2}$ and
$b_{pd} = 2.5 \pm 1 \ \rm{GeV}^{-2}$), the
proton dissociative to elastic production 
cross section ratio is  changed  from 0.75 to 0.5 and to 1.0, 
and the mass and width of the \rhoprim\ meson are varied: 
$M_{\rho^\prime} = 1450 \pm 150$~\mev\ and $ \Gamma_{\rho^\prime} = 
300~\pm 150$ \mev.

\item {\bf Uncertainties affecting {\boldmath $\rho$} production} \\
%
For the simulation of \rh\ meson production, the
$t$ slopes ($b_{el} = 7 \pm 1 \ \rm{GeV}^{-2}$ and
$b_{pd} = 1.7^{+0.8}_{-0.7} \ \rm{GeV}^{-2}$) and the proton 
dissociative to elastic production cross section ratio 
(0.75 $\pm$ 0.25) are changed, 
and the cross section dependences on $Q^2$ and $W$ are varied within
limits of the measurements in~\cite{h1-rho}.   
Furthermore, the $M_Y^2$ spectrum as implemented in DIFFVM is varied
from $1/M_Y^{2.15}$ to $1/M_Y^{1.85}$ and to $1/M_Y^{2.45}$.

\item {\bf Uncertainties in the detector response} \\
%
The energy threshold for the detection of LAr clusters not associated 
to tracks is varied between 300~MeV and
500~MeV;
the efficiencies of the PRT and FMD are modified within
experimental errors; the measurement of the polar angle of 
the scattered electron is changed by $\pm 0.5 \ {\rm mrad}$ and
the uncertainties in the trigger and the tracker efficiencies 
are included. 

\end{itemize}

For the measurements in section~\ref{sect:tdepend}, the dominant
systematic error is due to the uncertainty in the shapes 
of the backgrounds.


\section{\boldmath{$t^\prime$} dependences of spin
 density matrix elements}
                                  \label{sect:tdepend}

\subsection{\boldmath{$\Phi$} distributions and determination of
 (\boldmath{\rfivecomb}) and (\boldmath{\ronecomb})}
  \label{sect:combin}

\begin{figure}[htbp]
\vspace{-0.cm}
\begin{center}
\epsfig{file=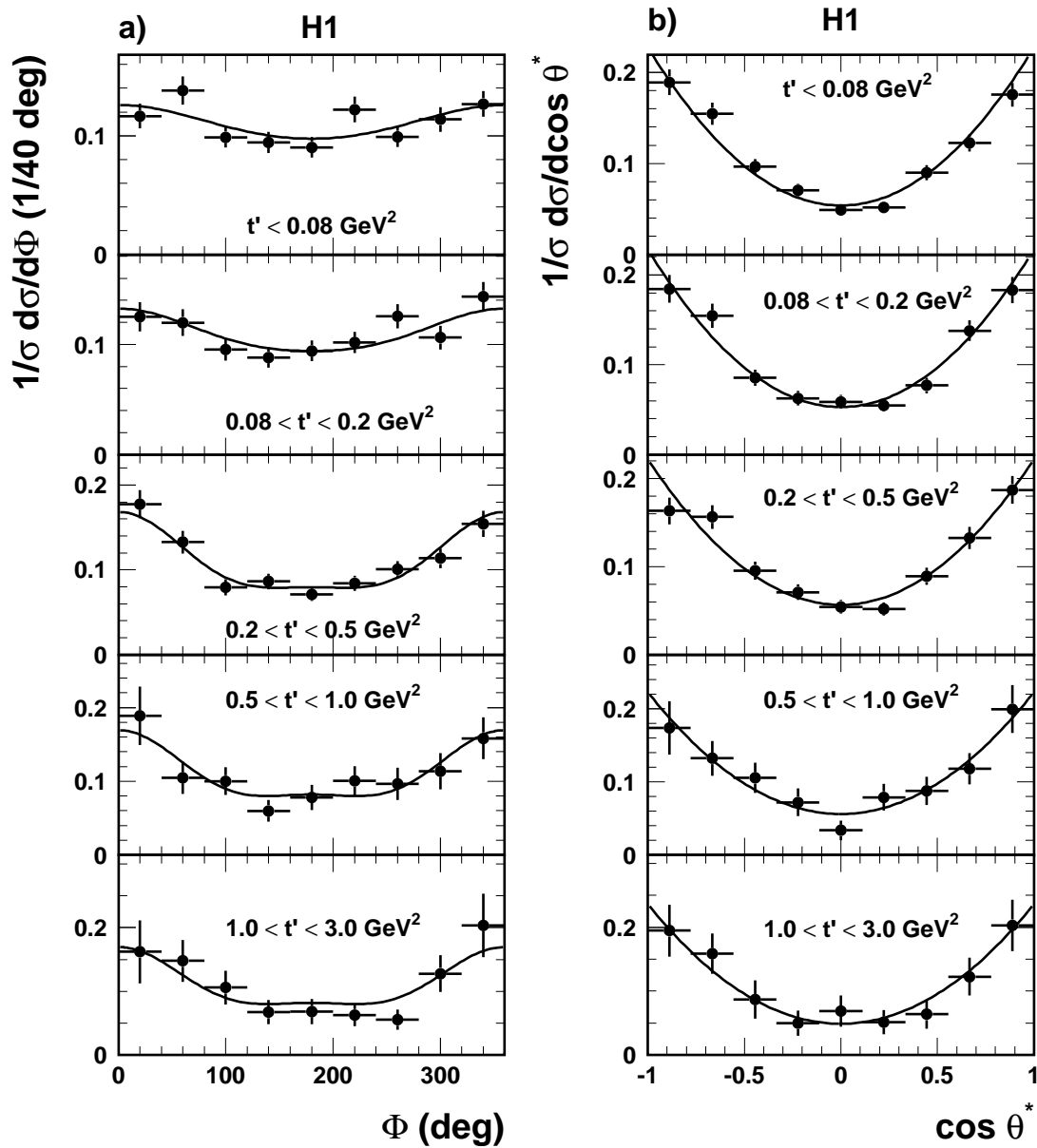,width=16.cm}
\end{center}
\vspace{-1.cm}
\caption{Normalised cross sections for
  \rh\ electroproduction in five bins in \tprim. The
  superimposed curves show the results of fits of a) 
  relation~(\protect\ref{eq:Phi}) and
  b) relation~(\protect\ref{eq:cost}). 
  The error bars represent the statistical errors.}
\label{fig:final}
\end{figure}

\begin{figure}[htbp]
\begin{center}
\vspace{-0.cm}
\epsfig{file=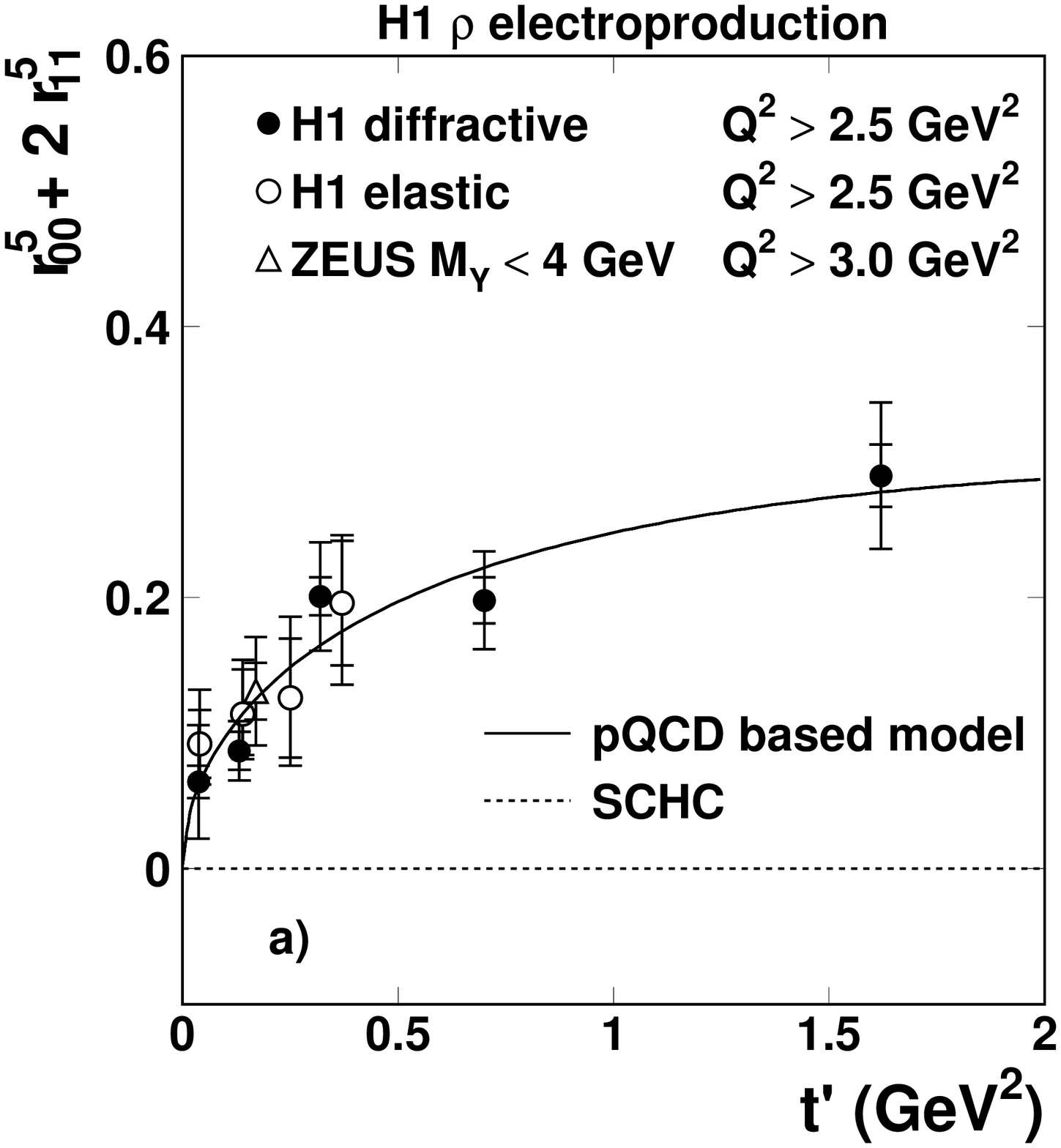,width=7.9cm} 
\epsfig{file=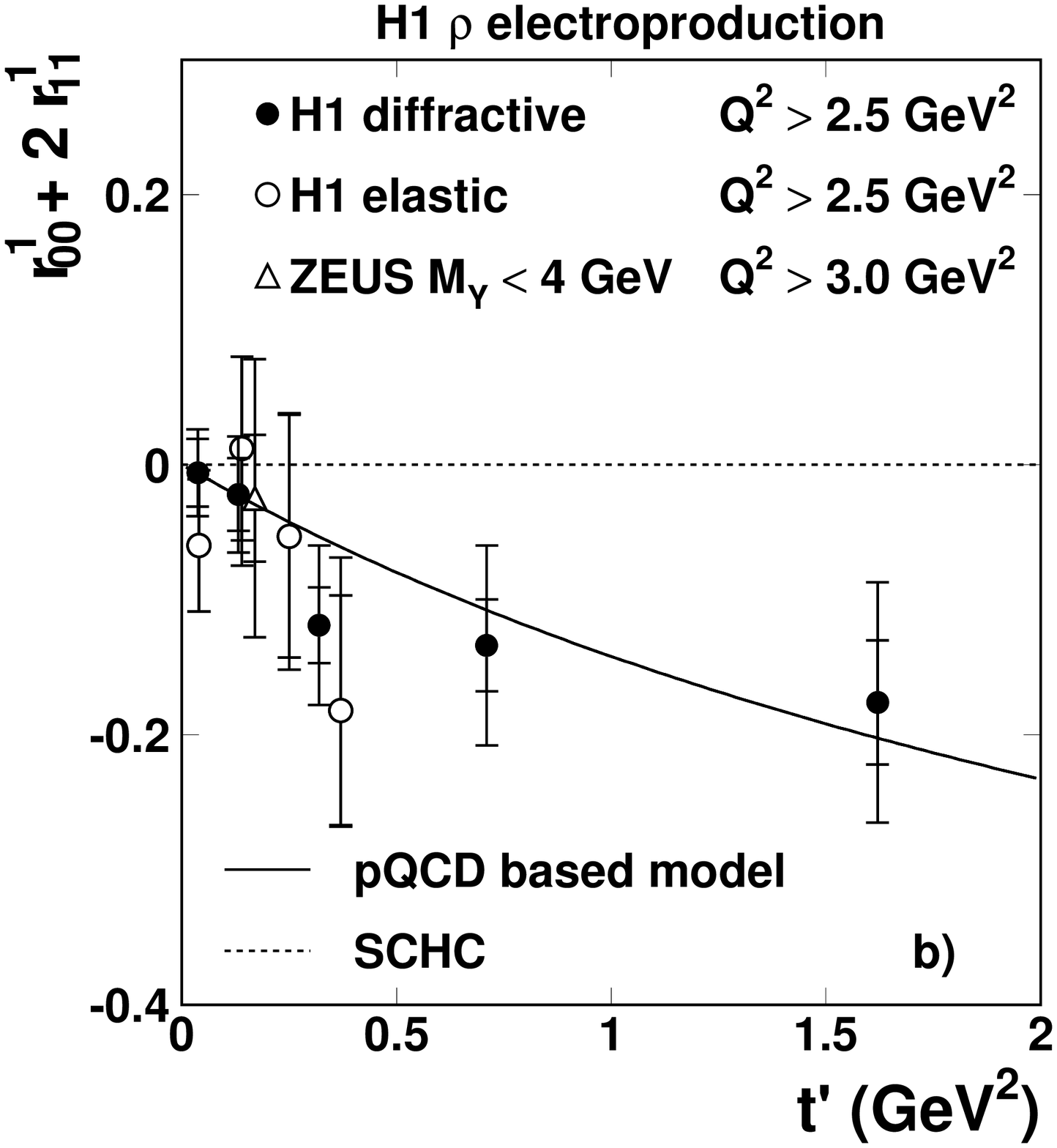,width=7.9cm}
\epsfig{file=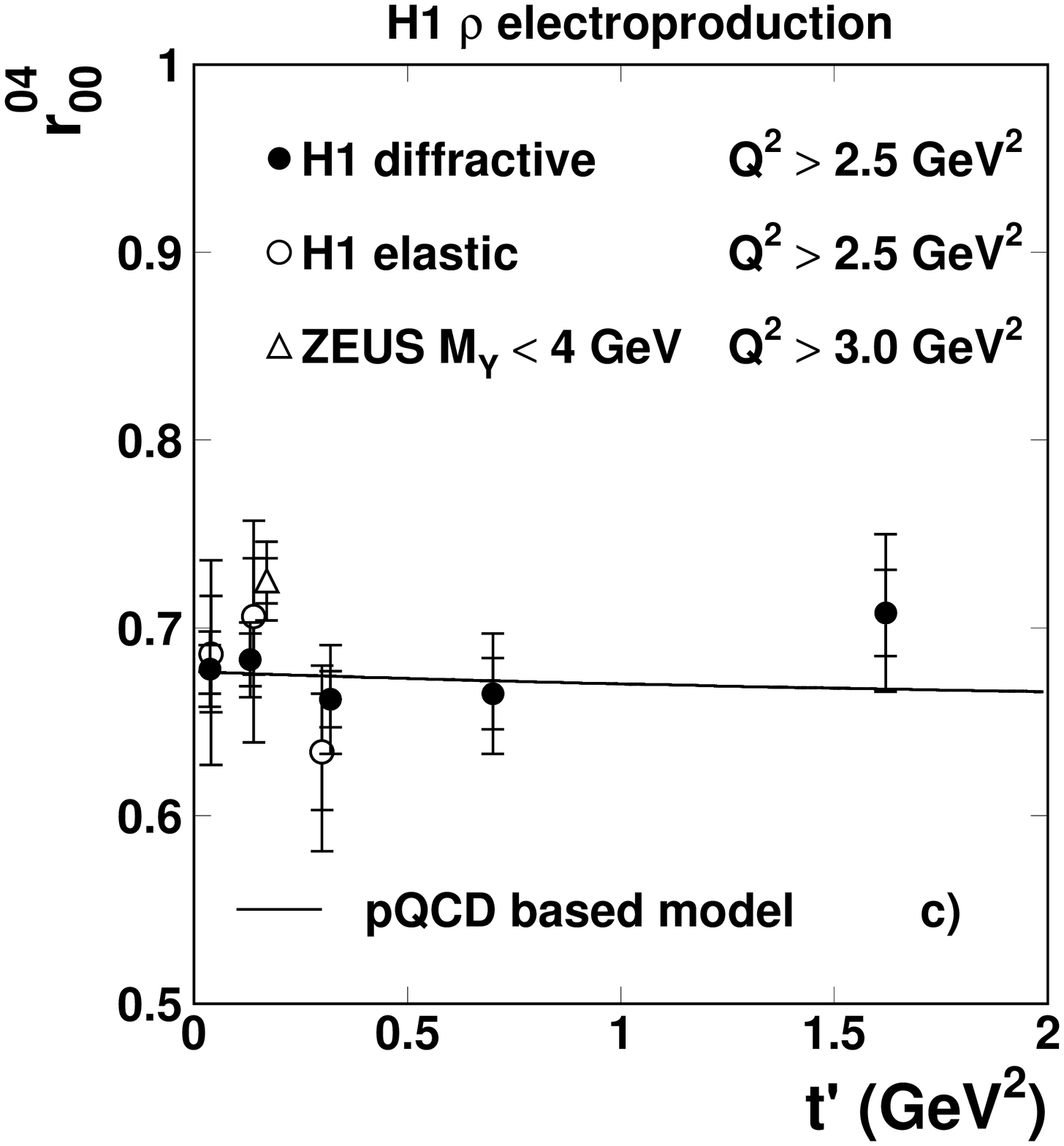,width=7.5cm}
\end{center}
\caption{Measurement of a) \protect\rfivecomb\ ,
b) \protect\ronecomb\ , 
c) \protect\rfour\ as a function of
\protect\tprim, for the present analysis and for
refs.~\protect\cite{h1-rho} (labeled ``H1 elastic'') 
and~\protect\cite{zeus}
(labeled ``ZEUS $M_Y < 4$ GeV''). \protect \nolinebreak \protect
The \protect \linebreak \protect
inner error bars represent the statistical errors,
the full error bars include the systematic errors added in 
quadrature.
The full curves correspond to the predictions of
the model~\protect\cite{ivanov} with parameters
extracted from the fit described in 
section~\ref{sect:global}.
The dashed lines in a) and b) correspond to SCHC.} 
\label{fig:dme1}
\end{figure}

Measurements of the \dme\ combinations \rfivecomb\ and \ronecomb\
are obtained from fits of eq.~(\ref{eq:Phi}) to the $\Phi$ 
distributions, fully corrected for the presence of backgrounds and
for acceptance, smearing and QED radiative effects, in the five 
\tprim\ bins shown in Fig.~\ref{fig:final}a.
For $0.5 < t^\prime < 3.0 \ \rm{GeV}^2$, only the tag sample is used,
in view of the much larger background in the notag sample 
(compare Figs.~\ref{fig:mass}c-d).
Given the small backgrounds (Figs.~\ref{fig:mass}a-b) and in order to
improve the statistical precision of the measurement, the tag and 
the notag samples are merged for \tprim\ $< 0.5 \ {\rm GeV^2}$. 
The measurements are given in Table~\ref{tab:values} and 
presented in Figs.~\ref{fig:dme1}a and~\ref{fig:dme1}b,
together with previous measurements for \tprim\ $< 0.5 \ \rm{GeV}^2$
in similar $W$ and \qsq\ ranges 
($Q^2 > 2.5 \ \rm{GeV}^2$ from ref.~\cite{h1-rho} 
and $Q^2 > 3.0 \ \rm{GeV}^2$ from ref.~\cite{zeus}).

\begin{table}[htbp]
\begin{center}
\begin{tabular}{|c|c|c|ccc|}
\hline
\hline
 Element & $t^\prime \ ({\rm GeV^{2}})$ 
 & $\langle t^\prime \rangle \ ({\rm GeV^{2}})$ & &  Measurement & \\
\hline
\hline
$  $ & $   t^\prime < 0.08 $ & 0.037 
   & 0.064 & $\pm$ 0.012 & $\pm$ 0.040 \\
$  $ & $  0.08 < t^\prime < 0.2 $ & 0.132 
   & 0.087 & $\pm$ 0.014 & $\pm$ 0.017 \\
$  r^{5}_{00} + 2r^{5}_{11}$ & $  0.2 < t^\prime < 0.5 $ & 0.320
   & 0.201 & $\pm$ 0.014 & $\pm$ 0.037 \\
$  $ & $  0.5 < t^\prime < 1.0 $ & 0.700
   & 0.198 & $\pm$ 0.017 & $\pm$ 0.032 \\
$  $ & $  1.0 < t^\prime < 3.0 $ & 1.620
   & 0.290 & $\pm$ 0.023 & $\pm$ 0.049 \\
\hline
$  $ & $ t^\prime < 0.08 $ & 0.037 
   &-0.006 & $\pm$ 0.025 & $\pm$ 0.020 \\
$  $ & $  0.08 < t^\prime < 0.2 $ & 0.132
   &-0.022 & $\pm$ 0.027 & $\pm$ 0.034 \\
$  r^{1}_{00} + 2r^{1}_{11}$ & $  0.2 < t^\prime < 0.5 $ & 0.320
   &-0.119 & $\pm$ 0.028 & $\pm$ 0.052 \\
$  $ & $  0.5 < t^\prime < 1.0 $ & 0.700
   &-0.134 & $\pm$ 0.034 & $\pm$ 0.066 \\
$  $ & $  1.0 < t^\prime < 3.0 $ & 1.620
   &-0.176 & $\pm$ 0.046 & $\pm$ 0.076 \\
\hline
$  $ & $ t^\prime < 0.08 $ & 0.037 
   & 0.678 & $\pm$ 0.013 & $\pm$ 0.015 \\
$  $ & $ 0.08 < t^\prime < 0.2 $ & 0.132 
   & 0.683 & $\pm$ 0.014 & $\pm$ 0.014 \\
$  r^{04}_{00}$ & $ 0.2 < t^\prime < 0.5 $ & 0.320
   & 0.662 & $\pm$ 0.015 & $\pm$ 0.025 \\
$  $ & $ 0.5 < t^\prime < 1.0 $ & 0.700
   & 0.665 & $\pm$ 0.019 & $\pm$ 0.026 \\
$  $ & $ 1.0 < t^\prime < 3.0 $ & 1.620
   & 0.708 & $\pm$ 0.023 & $\pm$ 0.035 \\
\hline
\hline
\end{tabular}
\end{center}
\caption{Measurement of the spin density matrix element combinations
 \rfivecomb , \ronecomb\ and \rfour\ in five bins of \tprim .
 The first errors are statistical, the second systematic.}
\label{tab:values}
\end{table}



Significant helicity non-conservation is observed in 
Fig.~\ref{fig:dme1}a for the combination \rfivecomb\
(SCHC would imply a zero value of the combination).
The \rfive\ matrix element is proportional (see relations~(\ref{eq:ampli})) 
to the product of the dominant non-flip amplitude 
$T_{00}$ and the $T_{01}$ amplitude, expected to be the largest 
helicity flip amplitude (see relations~(\ref{eq:hierarchy})).
In contrast, the $r^5_{11}$ matrix element has a contribution from the 
product of the non-dominant non-flip amplitude $T_{11}$ and the 
non-dominant single flip amplitude $T_{10}$, and a contribution 
dependent on the double flip amplitude.
The strong \tprim\ dependence of the \rfivecomb\ combination
is thus attributed mainly to the predicted~\cite{royen,niko,ivanov}
$\sqrt {\tprim}$ dependence of the ratio of 
the $T_{01}$ to the non-flip amplitudes.
Note that the \tprim\ dependence of the \rfivecomb\ combination is 
not exactly $\propto \sqrt {\tprim}$, as expected for the single-flip 
to the non-flip amplitude ratio, since it also involves the effect 
of the single and double-flip amplitudes
in the denominator $N$ of relations (\ref{eq:ampli}).

The values for \ronecomb\ are shown in Fig.~\ref{fig:dme1}b. They
are significantly different from zero 
and negative, which implies violation of SCHC.
This is the first observation of a significant
non-zero value of the \ronecomb\ \dme\ combination. 
The $r^1_{00}$ element, which gives a
negative contribution (see relations~(\ref{eq:ampli})), is proportional to 
the square of the single flip amplitude $T_{01}$.
The $r^1_{11}$ element is proportional to the product of the 
non-dominant non-flip amplitude $T_{11}$ and the double flip 
amplitude.
The sign of the combination thus gives information on the
relative strength of the $T_{01}T_{01}^\dagger$ and 
$T_{11}T_{1-1}^\dagger$ products of amplitudes.
It confirms that the $T_{01}$ amplitude 
is significantly larger than the double flip amplitude in
the present kinematical domain. 
The \tprim\ dependence of the \ronecomb\ combination is expected 
to be linear, up to effects of the single and double-flip amplitudes 
in the denominator $N$ of relations (\ref{eq:ampli}).


\subsection{\boldmath{\cost} distributions and 
       determination of \boldmath{\rfour}}
  \label{sect:rfour}

       
The \rfour\ spin density matrix element is extracted from the 
\cost\ distribution using relation~(\ref{eq:cost}).
Fig.~\ref{fig:final}b presents the fully 
corrected \cost\ distributions for five bins in \tprim.
The normalisations of the $\omega$, $\phi$ and \rhoprim\ backgrounds 
are estimated as described above.
The shape of the background is determined from the data
and found to be flat.
This is done by comparing the \cost\ distributions for the events 
with $\zeta < 0$ and $\zeta > 0$, which differ in the data whereas 
they are predicted to be similar by the \rh\ Monte Carlo simulation.
The difference is attributed to background. 

The extracted values of \rfour\ are presented in 
Table~\ref{tab:values} 
and in Fig.~\ref{fig:dme1}c, together with previous measurements
at low \tprim\ values in the same \qsq\ and $W$ range.
No significant variation of \rfour\ with \tprim\ is observed. 
This is expected from relation~(\ref{eq:ampli}) in view of the 
predicted \tprim\ independence of the ratio $|T_{11}|/|T_{00}|$, with 
small corrections from the other amplitudes.
This observation implies that the slopes of the 
exponentially falling $t$ distributions for the transverse and 
longitudinal $s$-channel helicity conserving amplitudes, 
$T_{00}$ and $T_{11}$, are very similar.


\section{QCD description of the measurements}
                                     \label{sect:global}

Perturbative QCD calculations for vector meson electroproduction
assume the factorisation of the non-perturbative from the perturbative 
contributions to the amplitudes.
Collinear factorisation has been demonstrated for longitudinal 
photons~\cite{cfs},
%
%
but logarithmic singularities are manifest for the 
transverse photon polarisation when the fraction $z$ of the 
longitudinal momentum carried by the quark approaches 0 or~1.
%
%
Non-perturbative effects were suggested to damp these 
singularities~\cite{brodsky}, in which case perturbative calculations 
become problematic.
However, as noted in~\cite{mrt}, these contributions cannot be 
large,\footnote
{Large non-perturbative contributions to the transverse 
amplitudes would imply a $t$ slope similar to the non-perturbative 
photoproduction case, $b_T \simeq 10$ \gevsqm , 
significantly larger than for the longitudinal cross section 
($b_L \simeq 6$ \gevsqm ). This is inconsistent with the very weak 
$t$ dependence of the \rfour\ matrix element (section~\ref{sect:rfour}).} 
and in the 
models~\cite{royen,niko,ivanov} pQCD is expected 
to be valid also for the transverse amplitudes.

In the models~\cite{royen,niko,ivanov} all amplitudes are proportional, 
in the leading $\log (Q^2)$ approximation, to the gluon density 
in the proton (except for non-perturbative contributions in the 
double-flip amplitude).\footnote{For large 
mass vector mesons and/or large \qsq , skewed parton distributions 
should be used.}
More specifically, in the model of Ivanov and Kirschner~\cite{ivanov},
the gluon distribution in the proton $xG(x,\tilde{Q}^2)$ is probed 
at the hard scale $\tilde{Q}^2 = z(1-z)Q^2 \leq\ Q^2/4$.
Following a suggestion of Martin, Ryskin and Teubner~\cite{mrt},
the scale dependence of the gluon density in the leading 
log $\tilde{Q}^2$ approximation is parameterised as
$G(x, \tilde{Q}^2) = G(x,Q_0^2) \ [\tilde{Q}^2 / Q_0^2]^\gamma$, 
where the gluon anomalous dimension $\gamma$ is taken as \qsq\ 
independent.
This permits the singularities as $z \rightarrow 0,1$
to be avoided and 
factorisation is effectively (but not necessarily exactly) restored 
for the transverse amplitudes.

The absolute values and the \qsq\ and $t$ dependences of the 
four independent ratios\footnote{Under the assumptions 
of natural parity exchange and of purely imaginary amplitudes.} 
of the amplitudes $T_{\lambda_{V \! M}\lambda_{\gamma}}$  
are predicted by the model~\cite{ivanov} with two independent 
parameters: the effective gluon anomalous dimension $\gamma$, and 
the effective mass $m$ of the incoming $q \bar q$ pair.
These two free parameters are obtained from a fit to the \tprim\ 
dependence of the 15 measurements of spin density 
matrix elements in the present analysis.\footnote
{The amplitude $T_{1-1}$, which 
contains a non-perturbative part in \cite{ivanov}
and is expected to be very small 
(see relations~(\ref{eq:hierarchy})), can be set to zero or included as a free 
parameter in the fit without affecting the results.} 
The fit gives an excellent description of the data: 
$\chi^2/ndf = 0.41$ when the full errors are used 
and $\chi^2/ndf = 1.71$ for 
statistical errors only.
The fitted values of the parameters are
$ \gamma = 0.60 \pm 0.09$ and $m = 0.58 \pm 0.04 $~GeV.
The errors are the quadratic combination of the statistical and 
systematic errors, the latter obtained by repeating the fits with
the data shifted by each
systematic uncertainty in turn.
The dominant error comes from the uncertainty in the background 
shape.
The results of the fit are shown as solid lines in 
Figs.~\ref{fig:dme1}a-c.
As can be observed, the 11 low \tprim\ measurements of ref.~\cite{h1-rho}, 
which correspond to the same $Q^2$ and $W$ ranges as the present data, 
are also very well described; their inclusion in the 
fit does not change the quantitative results significantly.

According to the parton distributions in PDFLIB~\cite{pdflib}, the
extracted value of $\gamma$ corresponds to the \qsq\ evolution of the
gluon density for $\tilde{Q}^2 \simeq 5.0 $ \gevsq\ which is much higher 
than the average $\langle \tilde{Q}^2 \rangle \leq\ \langle Q^2 /4
\rangle \simeq\ 1.3$ \gevsq\ in the data.
The $\gamma$ parameter was introduced in the model
to restore factorisation which is otherwise broken by end-point effects 
($z \rightarrow 0,1$) in the transverse amplitudes. 
It does not need to be strictly interpreted as describing the 
evolution of the gluon distribution in the proton 
at the specified $\tilde{Q}^2$~\cite{rk}.
On the other hand, the disagreement may suggest that
the model does not apply in the full \qsq\
range of the present data.
The parameter $m = 0.58 \pm 0.04$~GeV is slightly below the \rh\ 
meson mass but belongs to a mass range where the quark pair is highly
likely to recombine into a \rh\ meson 
(cf the parton-hadron duality arguments in~\cite{mrt}).

\section{Conclusions}
                                     \label{sect:concl}

A measurement has been performed of \rh\ meson diffractive
electroproduction in the range $2.5  <  Q^2  <  60~{\rm GeV^{2}}$, 
$40  <  W  < 120~{\rm GeV}$ and $0  <  t^\prime  < 3~{\rm 
GeV^{2}}$.
The \rfour\ \dme\ and the combinations \rfivecomb\ and
\ronecomb\ have been measured as functions of \tprim.

No significant \tprim\ dependence is observed for \rfour .
A significant violation of {\it s}-channel helicity conservation,
increasing with \tprim ,
is observed in the \rfivecomb\ combination.
It is consistent with a $\sqrt {\tprim}$ dependence of the ratio 
of the amplitude $T_{01}$ to the non-flip 
amplitudes, $T_{01}$ being the amplitude for the transition 
from a transverse photon to a longitudinal \rh\ meson.
The \ronecomb\ combination is different from zero 
and negative;
this is the first observation of a significant non-zero value of
this combination. 
The sign gives information on the
relative strength of the $T_{01}T_{01}^\dagger$ and 
$T_{11}T_{1-1}^\dagger$ amplitude products.
Together with the \rfivecomb\ measurement, it confirms that 
the $T_{01}$ amplitude is relatively large in the present kinematical 
domain, and significantly larger than the double flip amplitude. 

A fit of the pQCD model of Ivanov and
Kirschner~\cite{ivanov} to the present 15 measurements of \dmes\ 
gives a good description of the \tprim\ dependence of the data.
The value $\gamma = 0.60 \pm 0.09$ is obtained for the effective 
parameter describing the \qsq\ dependence of the gluon density and
$m = 0.58 \pm 0.04$~GeV is extracted for the average effective mass
of the incoming $q \bar q$ pair. Thus the data are broadly compatible
with a diffractive mechanism based on the exchange of two gluons, with
non-conservation of $s$-channel helicity occuring only when the
photon longitudinal momentum is shared asymmetrically between
the quark and the antiquark~\cite{royen,niko,ivanov}.

\section*{Acknowledgements}

We are grateful to the HERA machine group whose outstanding
efforts have made and continue to make this experiment possible. 
We thank the engineers and technicians for their work in 
constructing and now maintaining the H1 detector, our funding 
agencies for financial support, the DESY technical staff for 
continual assistance, and the DESY directorate for the hospitality 
which they extend to the non DESY members of the collaboration.
We also thank M.~Diehl, D.Yu.~Ivanov, I.P.~Ivanov and 
R.~Kirschner for helpful contributions.

\end{document}